\documentclass{article}
\usepackage{spconf,amsmath,graphicx,tikz,booktabs}
\usepackage{float}
\usepackage[colorlinks=true]{hyperref}
\usepackage{fancyhdr}
\usetikzlibrary{positioning}
\usetikzlibrary{calc}
\usetikzlibrary{shapes.multipart}

\fancypagestyle{firstpage}{%
\fancyhf{}
\fancyfoot[C]{\copyright2018 IEEE. Published in the IEEE 2018 International Conference on Acoustics, Speech, and Signal Processing (ICASSP 2018), scheduled for 15-20 April 2018 in Calgary, Alberta, Canada.}

}
\makeatletter
\newcommand{\doubletilde}[1]{{%
  \mathpalette\double@tilde{#1}%
}}
\newcommand{\double@tilde}[2]{%
  \sbox\z@{$\m@th#1\tilde{#2}$}%
  \ht\z@=.9\ht\z@
  \tilde{\box\z@}%
}
\makeatother
\title{Spectral feature mapping with mimic loss \\ for robust speech recognition}
\name{Deblin Bagchi \qquad Peter Plantinga \qquad Adam Stiff \qquad Eric Fosler-Lussier}
\address{Department of Computer Science and Engineering \\
The Ohio State University, Columbus, OH, USA}
\begin{document}
%
\maketitle
\thispagestyle{firstpage}
\newpage
\begin{abstract}

For the task of speech enhancement, local learning objectives are agnostic to phonetic structures helpful for speech recognition. 
We propose to add a global criterion to ensure de-noised speech is useful for downstream tasks like ASR. We first train a spectral classifier on clean speech to predict senone labels. Then, the spectral classifier is joined with our speech enhancer as a noisy speech recognizer. This model is taught to imitate the output of the spectral classifier alone on clean speech. This \textit{mimic loss} is combined with the traditional local criterion to train the speech enhancer to produce de-noised speech. Feeding the de-noised speech to an off-the-shelf Kaldi training recipe for the CHiME-2 corpus shows significant improvements in WER.
\end{abstract}
\section{Introduction}
\label{sec:intro}
Automatic Speech Recognition (ASR) has shown tremendous progress over the years in recognizing clean speech. However, traditional DNN-HMM ASR systems still suffer from performance degradation in the presence of acoustic interference, such as additive noise and room reverberation.  Some strategies for building a noise-robust speech recognizer include using noise-invariant features \cite{wang2016joint}, augmented data \cite{du2016ustc}, bulkier acoustic models like LSTMs and CNNs \cite{du2016ustc,yoshioka2015ntt} and sophisticated language models \cite{du2016ustc,yoshioka2015ntt}. Few groups, however, have looked at systems that train only a speech enhancement model, which can be used for different tasks.


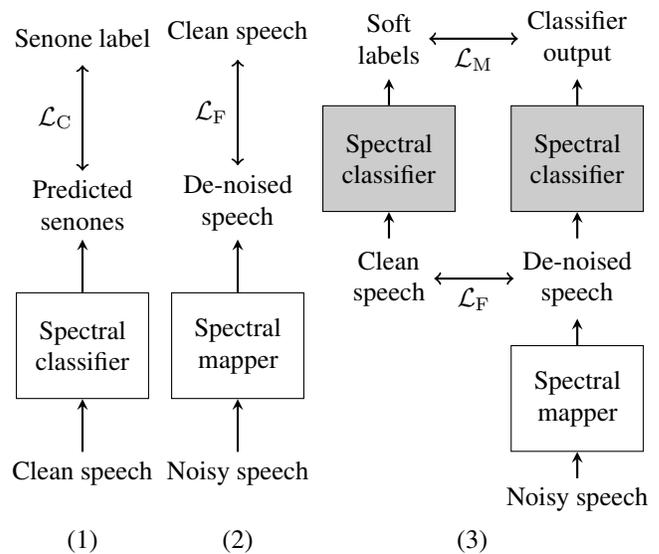
\begin{figure}[b!]
\vspace*{-0.2in}
\hrulefill
\label{fig:systemdesc}
\tikzstyle{model} = [rectangle, draw, minimum height=4em, minimum width=5em,align=center]
\begin{tikzpicture}
	\node (a) at (0,0) {(1)};
	\node[above=1em of a] (crit_in) {Clean speech};
    \node[model, above=2em of crit_in] (crit) {Spectral \\ classifier};
    \node[above=2em of crit,align=center,minimum height=2em] (crit_out) {Predicted \\ senones};
    \node[above=4em of crit_out, minimum height=2em] (senone) {Senone label};
    
    \draw[-stealth,thick] (crit_in) -- (crit);
    \draw[-stealth,thick] (crit) -- (crit_out);
    \draw[-stealth,thick,<->] (crit_out) -- node[left,align=center] {$\mathcal{L}_{\rm C}$} (senone);
    
    \node[right=4em of a] (b) {(2)};
    \node[above=1em of b] (act_in) {Noisy speech};
    \node[model, above=2em of act_in] (act) {Spectral \\ mapper};
    \node[above=2em of act,align=center,minimum height=2em] (act_out) {De-noised \\ speech};
    \node[above=4em of act_out, minimum height=2em] (clean) {Clean speech};

	\draw[-stealth,thick] (act_in) -- (act);
    \draw[-stealth,thick] (act) -- (act_out);
    \draw[-stealth,thick,<->] (act_out) -- node[left,align=center] {$\mathcal{L}_{\rm F}$} (clean);
    
    \node[right=7em of b] (c) {(3)};
    
    \node[above right=0em and 0em of c] (joint_noisy) {Noisy speech};
    \node[model, above=1em of joint_noisy] (joint_actor) {Spectral \\ mapper};
    \node[above=1em of joint_actor,align=center,minimum height=3em] (joint_act_out) {De-noised \\ speech};
    \node[left=3em of joint_act_out,align=center, minimum height=3em] (joint_clean) {Clean \\ speech};
    
    \node[model, above=1em of joint_clean, fill=gray!40] (output_critic) {Spectral \\ classifier};
    \node[above=1em of output_critic,align=center,minimum height=3em] (soft_labels) {Soft \\ labels};
    
    \node[model, above=1em of joint_act_out, fill=gray!40] (joint_critic) {Spectral \\ classifier};
    \node[above=1em of joint_critic,align=center,minimum height=3em] (joint_crit_out) {Classifier \\ output};
    \draw[-stealth,thick] (joint_noisy) -- (joint_actor);
    \draw[-stealth,thick,<->] (joint_act_out) -- node[below] {$\mathcal{L}_{\rm F}$} (joint_clean);
    \draw[-stealth,thick] (joint_clean) -- (output_critic);
    \draw[-stealth,thick] (output_critic) -- (soft_labels);
    \draw[-stealth,thick] (joint_actor) -- (joint_act_out);
    \draw[-stealth,thick] (joint_act_out) -- (joint_critic);
    \draw[-stealth,thick] (joint_critic) -- (joint_crit_out);
    \draw[-stealth,thick,<->] (joint_crit_out) -- node[below] {$\mathcal{L}_{\rm M}$} (soft_labels);

\end{tikzpicture}
\caption{Our speech enhancement system is trained in three steps. (1) The spectral classifier is trained to predict senone labels with cross-entropy criterion (classification loss, $\mathcal{L}_{\rm C}$). (2) The spectral mapper is pre-trained to map from noisy speech to clean speech using MSE criterion (fidelity loss, $\mathcal{L}_{\rm F}$). (3) The spectral mapper is trained using joint loss from both the clean speech and the outputs of the classifier when fed clean speech (mimic loss, $\mathcal{L}_{\rm M}$). The gray models have frozen weights.}
\end{figure}

A speech enhancement front-end refers to a performance-boosting denoising technique that can be attached to any standard automatic speech recognition model. Some deep learning models for speech enhancement attempt to estimate an ideal ratio mask (IRM) for removing noise from a speech signal \cite{narayanan2015improving}. Others utilize spectral mapping in the signal domain 
\cite{han2014learning,han2015learning} or in the feature domain 
\cite{han2015deep,bagchi2015combining} to directly predict features.

Recent work in computer vision has seen notable success in addressing the problem of poor resolution in modified input data, using a framework broadly referred to as Generative Adversarial Networks (GANs). Fundamentally, these gains are achieved by the injection of auxiliary or proxy learning objectives into more traditional pipelines. These auxiliary objectives exploit an adversarial relationship between two neural networks (a \textit{generator} and a \textit{discriminator}) to find a Nash equilibrium in which generating sensible data is the optimal behavior for the generator \cite{isola2016image}. This has the effect of refining the distribution of the generated data closer to some desirable outcome relative to a system trained without the auxiliary objective, e.g., sharper, more realistic generated images.

In light of these successes, in particular that of \cite{isola2016image},  inserting an auxiliary \textit{realism} objective into a speech denoising  pipeline seems to be a natural avenue to pursue improved performance.  However, initial experiments with GANs conditioned on noisy speech inputs, in which the discriminator network was trained to distinguish between real and fake clean/noisy input pairs, exposed the well-known mode collapse problem endemic to GANs \cite{salimans2016improved}. Results failed to improve upon simple baselines of noisy-to-clean speech mappings trained with only MSE loss. Other attempts have also failed to improve upon a DNN baseline, as in \cite{michelsanti2017conditional}. We speculate that the difference in performance from experiments in the visual domain may be due to the relatively non-local structure of the speech signal in the frequency axis (i.e. harmonic structure), as well as the smoothness of speech features as compared to images.

The work described in this paper is motivated by the observations of over-simple
outputs from GANs that seemed to be stuck in mode collapse orbits. We hypothesize that a training objective that can provide stronger feedback than a simple real or fake determination will be better able to guide the parameters of the denoising network towards producing output that behaves like actual speech. While our resulting system retains none of the properties of being generative or trained adversarially, the insight to use an auxiliary task to improve the denoising process is drawn from that body of work.

The auxiliary global objective that we add to our local criterion is the behavioral loss of a classifier trained on clean speech. We call this additional objective the \textit{mimic loss}. First, we train a senone classifier using clean speech as input,  and a spectral mapper network \cite{bagchi2015combining,han2015learning} using parallel noisy and clean speech frame pairs. Next, we freeze the weights of the acoustic model and join our pre-trained spectral mapper to it. We then pass noisy speech frames to train our spectral mapper with a joint objective, i.e. a weighted sum of the traditional fidelity loss and mimic loss. The mimic loss, then, is MSE loss with respect to the softmax (or pre-softmax) outputs of the classifier fed with the corresponding clean speech frame. See figure \ref{fig:systemdesc} for a graphical depiction of the model.

This technique of using one model to teach another was proposed by Ba and Caruna \cite{ba2014deep} for the task of model compression. In their work, they introduce student-teacher learning, where separate teacher and student models are trained to do the same task. Mimic loss, on the other hand, is used to train the student model to do a different task from the teacher model.

\section{prior work}
\label{sec:prior_work}
To deal with noise, many DNN based methods have been proposed to improve the robustness of ASR systems. In acoustic modeling, using Convolutional Neural Networks (CNNs), such as in \cite{qian2016very} and Long Short Term Memory Networks (LSTMs) in \cite{chen2015integration} has resulted in an improvement in performance. LSTMs have also been successfully used as speech enhancement front-ends in \cite{weninger2015speech,chen2015integration}. 

Spectral mapping has been used to generate clean speech signals. However, in \cite{bagchi2015combining,han2015deep} they use only a local learning objective. Student-teacher networks have been used to improve the quality of noisy speech recognition \cite{markov2016robust,watanabe2017student,li2017large}. Our model uses mimic loss instead of student-teacher learning, which means the speech enhancer is not jointly trained with a particular acoustic model. This speech enhancer could be used as a pre-processor for any ASR system, or for another similar dataset. This modularity is the strength of mimic loss.

\section{system description}
\label{sec:sysdesc}

In this section, we will describe the major components of our system: namely, 
the spectral mapper, the spectral classifier, and the overall framework binding the two together.

\subsection{Spectral mapping}
\label{ssec:specmap}
Spectral mapping improves performance of the speech recognizer by learning a mapping from noisy spectral patterns to clean features. We train a DNN-based spectral mapper for feature denoising. In our previous work \cite{han2015deep,bagchi2015combining}, we have shown that a DNN-based spectral mapper, which takes noisy spectrogram as input to predict clean filterbank features for ASR, yields good results on the CHiME-2 noisy and reverberant dataset.

Specifically, we first divide the input time-domain signals into 25-ms frames with a 10-ms frame shift, and then apply short time Fourier transform (STFT) to compute log spectral magnitudes in each time frame. For a 16 kHz signal, each frame contains 400 samples, and we use 512-point Fourier transform to compute the magnitudes, forming a 257-dimensional log magnitude vector. 
%
%
%
Each noisy spectral component $x^k_m$ for frequency $k$ at time slice $m$ is augmented on the input by the deltas and double deltas, as well as a five frame window (designated $\tilde{x}_m^k=[x_{m\pm 5}^k$]), leading to the dimensionality of $\tilde{x}_m$ being $257 \cdot 3 \cdot 11=8481$.  Similarly, we define $y_m$ to be the clean spectral slice at time $m$.



We then use a feed-forward neural network $f(\cdot)$ to map noisy spectral slices $\tilde{x}_m$ to clean spectral features $y_m$ using an MSE loss function, which we call {\em fidelity loss}.

\begin{equation}\label{eqn1}
\begin{split}
 \mathcal{L}_{\rm FIDELITY}(\tilde{x}_m,y_m) & = \frac{1}{K}\sum_{k=1}^{K} (y_m^k - f(\tilde{x}_m)^k)^{2}  \\
\end{split}
\end{equation}

\subsection{Spectral classifier}
\label{ssec:spec_clas}
The spectral classifier is similar to the traditional DNN acoustic model trained to classify a stacked clean spectral pattern $\tilde{y}_m$ to its corresponding senone class $z_m$. We train the classifier using a cross entropy criterion; critically, once the classifier is trained, we freeze the weights as a model of appropriate behavior under clean speech.




\subsection{Joint loss}
\label{ssec:jl}
We can define the {\em mimic loss} as the mean square difference between a $D$-dimensional representation $g$ within the spectral classifier evaluated on clean speech $y_m$ and its paired cleaned speech $f(\tilde{x}_m)$:

\begin{equation}\label{mimic_loss}
\begin{split}
 \mathcal{L}_{\rm MIMIC}(\doubletilde{x}_m, \tilde{y}_m) = \frac{1}{D}\sum_{d=1}^{D} (g(\tilde{y}_m)^d - g(\tilde{f}(\tilde{x}_m))^d)^2
\end{split}
\end{equation}

\noindent We experimented with two different representations for $g(\cdot)$: the posterior output of the senones after softmax normalization ({\em post-softmax}) and the layer outputs prior to the softmax normalization ({\em pre-softmax}).

While training the speech enhancer, we found that using only mimic loss was not enough to allow the model to converge. We speculate that the task of predicting senones is too different from the task of predicting clean speech for the error signal to drive the output of the speech enhancer to actually look like speech. 
%
Combining the fidelity and mimic losses into a joint loss allows the enhancer to better imitate the behavior of the classifier under clean speech while keeping the projection of noisy speech closer to clean speech.

\begin{equation}\label{joint_loss}
\begin{split}
 \mathcal{L}_{\rm JOINT} = \mathcal{L}_{\rm FIDELITY} + \alpha \mathcal{L}_{\rm MIMIC}
\end{split}
\end{equation}

\noindent The hyper-parameter $\alpha$ is used to ensure
that both of the losses that make up the joint loss have a similar magnitude. We used 0.1 for pre-softmax and 1000 for post-softmax.

\section{experimental setup}
\label{sec:exp_setup}
We evaluate the effectiveness of our proposed method on Track 2 of the CHiME-2 challenge \cite{vincent2013second}, which is a medium-vocabulary task for word recognition under reverberant and noisy environments without speaker movements. In this task, three types of data are provided based on the Wall Street Journal (WSJ0) 5k vocabulary read speech corpus: clean, reverberant and reverberant+noisy. The clean utterances are extracted from the WSJ0 database. The reverberant utterances are created by convolving the clean speech with binaural room impulse responses (BRIR) corresponding to a frontal position in a family living room. Real-world non-stationary noise background recorded in the same room is mixed with the reverberant utterances to form the reverberant+noisy set. The noise excerpts are selected such that the signal-to-noise ratio (SNR) ranges among -6, -3, 0, 3, 6 and 9 dB without scaling. The multi-condition training, development and test sets of the reverberant+noisy set contain 7138, 2454 and 1980 utterances respectively, which are the same utterances in the clean set but with reverberation and noise at 6 different SNR conditions.

Our system is monaural. In our experiments, we simply average the signals from the left and right ear. A GMM-HMM system is built using the Kaldi toolkit \cite{povey2011kaldi} on the clean utterances in the WSJ0-5k to get the senone state for each frame of the corresponding noisy-reverberant utterances. The initial clean alignments are obtained by performing forced alignment on the clean utterances. To refine the initial clean alignments, we further trained a DNN-based acoustic model using the filterbank features of the clean utterances, and re-generate clean alignments. These clean alignments are used as the labels for training all the acoustic models in this study. Note that the DNN-HMM hybrid system built on the clean utterances is a powerful recognizer. It achieves 2.3\% WER on the clean test set of the WSJ0-5k dataset.

\begin{table}[b]
\begin{center}
\addtolength{\tabcolsep}{-2px}
\begin{tabular}{lcc}
\toprule
Spectral input to Kaldi & CE WER & sMBR WER \\
\midrule
No enhancement & 18.0 & 17.3 \\
Enhancement via fidelity loss &  17.5 & 16.5  \\
Enhancement via joint loss \\
\qquad w/ post-softmax mimic loss  &  16.5 & 15.7  \\
\qquad w/ pre-softmax mimic loss & \textbf{15.7} & \textbf{14.7}\\
\bottomrule
\end{tabular}
\caption{Experimental results on the CHiME2 test set; CE WER is the word error rate of a DNN-HMM hybrid system trained using a cross-entropy criterion. sMBR WER is the error rate after sequential minimum Bayes risk training.}
\label{tab:results}
\end{center}
\end{table}

\begin{table}[t]
\begin{center}
\addtolength{\tabcolsep}{-2px}
\begin{tabular}{lcccccc}
\toprule
Enhancement &-6 dB & -3 dB & 0 dB & 3 dB & 6 dB & 9 dB \\
\midrule
None  & 29.8 & 22.3 & 17.6 & 13.4 & 10.9 & 9.7 \\
Fidelity loss & 29.3 & 20.6& 16.2 & 12.4 & 10.9 & 9.2 \\
Joint loss & \textbf{25.9} & \textbf{19.6} & \textbf{14.8} & \textbf{10.9} &\textbf{9.0} & \textbf{8.0} \\
\bottomrule
\end{tabular}
\addtolength{\tabcolsep}{2px}
\caption{Experimental results on the CHiME2 test set, broken down across different SNRs. Mimic loss based on pre-softmax units. Bold indicates the best performing system for each evaluation subset. }
\label{tab:sim}
\end{center}
\end{table}

\begin{table}[b]
\begin{center}
\begin{tabular}{lcccccccc}
\toprule
Study & WER \\
\midrule
Wang et.al \cite{wang2016joint}& 10.6 \\
Weninger et.al\cite{weninger2015speech} &  13.8  \\
\midrule
proposed approach & 14.7 \\
\bottomrule
Narayanan-Wang\cite{narayanan2015improving}  &  15.4 \\
Chen et. al \cite{chen2015integration}& 16.0\\
\bottomrule
\end{tabular}
\end{center}
\caption{Performance comparison with other studies on the CHiME2 test set.}
\label{tab:comparisons}
\hspace{1cm}
\end{table}

\subsection{Description of the acoustic model}

In order to determine the effectiveness of the additional criterion, we train a model using the denoised features with an off-the-shelf Kaldi recipe. The DNN-HMM hybrid system is trained using the clean WSJ0-5k alignments generated using the method stated above. The DNN has 7 hidden layers, with 2048 sigmoid neurons in each layer and a softmax output layer. Splicing context size for the filter-bank features was fixed at 11 frames, with the minibatch-size being 1024. After that, we train the DNN with sMBR sequence training to achieve better performance. We regenerate the lattices after the first iteration and train for 4 more iterations. We use the CMU pronunciation dictionary and the official 5k close-vocabulary tri-gram language model in our experiments. 

\subsection{Description of the spectral classifier}

The spectral classifier network is a multilayer feed-forward network which classifies clean speech frames as one of 1999 senone labels. We use 6 layers of 1024 neurons with Leaky ReLU activations and batch normalization between all the layers. While training the spectral classifier on clean speech, we apply softmax after the topmost layer, and use a cross entropy criterion to teach the network to produce senones.


\subsection{Description of the spectral mapper}

The spectral mapper is composed of a simple two-layer feed-forward network with 2048 neurons in each layer. The fact that this model is simple means it can be used in some low-resource situations, and is fast to use. Note that mimic loss can be applied to improve the results of any kind of speech enhancer. To regularize the network, we use batch normalization and a dropout rate of 0.5 between every layer. We also use ReLU activations after each layer. The classifier and mapper are written using TensorFlow\footnote{Code at \href{https://github.com/OSU-slatelab/actor_critic_specmap}{https://github.com/OSU-slatelab/actor\_critic\_specmap}}.

\subsection{Results}
\label{sec:res}

We see in Table~\ref{tab:results} that the outputs of the recognizer before the softmax provide a better target for the noisy speech recognizer, as suggested in \cite{ba2014deep}, even though our setup is quite different from theirs. The difference in performance between pre- and post-softmax targets may be due to a mismatch between target domain and loss criterion; ongoing work suggests that a cross-entropy mimic loss on post-softmax targets performs similarly to MSE on pre-softmax. Furthermore, the information loss in the softmax normalization may ``broaden" the allowable spectral mappings, harming generalization. This suggests that it is helpful for the noisy recognizer not only to have the same targets as the clean recognizer, but also to learn to behave in the same way as the clean recognizer. 

We also demonstrate this fact by training the noisy speech recognizer using hard targets rather than the soft targets of mimic loss. This caused the joint loss in the spectral mapper to diverge, providing more evidence that the noisy speech recognizer must learn to mimic the behavior, not just the targets of the clean speech recognizer.
Finally, in Table~\ref{tab:sim} we break down our results into different SNRs and see that the gains are consistent over all Signal-to-Noise levels.

\subsection{Comparison with other systems}
For context, we show in Table \ref{tab:comparisons} the performance of our system relative to other published results in the field.  The better-performing models in this list use more sophisticated models (like RNNs and LSTMs) for front-end speech enhancement \cite{chen2015integration,weninger2015speech} and acoustic modeling \cite{chen2015integration}, as well as noise-invariant features \cite{wang2016joint} (e.g., PNCC, MRCG). We use an off-the-shelf Kaldi recipe using DNN-HMM to do speech recognition, as well as a simple 2 layer feed-forward network to do spectral mapping. Again, mimic loss can theoretically be used to improve the results of any front-end system.

\section{conclusion}
\label{sec:conc}

We have proposed a speech enhancement criterion, called mimic loss, which can be used to produce speech that is useful for downstream ASR tasks. The mimic loss comes from comparing the outputs of a frozen clean speech recognizer, before softmax is applied, on clean and enhanced speech. 
This configuration allows the speech enhancement output to be used as clean speech for any downstream task. We see that with mimic loss, the spectral mapper learns to produce more detailed speech data, retaining features that fidelity loss alone fails to model.  Mimicking the behavior of the pre-softmax layer of the classifier was superior to mimicking the output of the senone posterior estimates; in general, the lower error rates show that these features are helpful for downstream tasks.

In future work, we propose to extend this work by matching every layer of the phone recognizer, rather than just the inputs and outputs. We could also use a variety of models for speech enhancement to demonstrate the effectiveness of mimic loss. Another avenue is to evaluate the output of our system in multiple domains to determine the effectiveness of this approach at learning domain-invariant representations of speech. Finally, we can train a more sophisticated acoustic model, rather than using an off-the-shelf Kaldi recipe.

\section{Acknowledgements}
This material is based upon work supported by the National Science Foundation under Grant IIS-1409431.  We also thank the Ohio Supercomputer Center (OSC) \cite{OhioSupercomputerCenter1987} for providing us with computational resources.
\bibliographystyle{IEEEbib}
\bibliography{strings,refs}

\end{document}